# Ultrafast Proton Delivery with Pin Ridge Filters (pRFs): A Novel Approach for Motion Management in Proton Therapy

Ahmal Jawad Zafar[1], Xiaofeng Yang[1], Yinan Wang[1], Zachary Diamond[1], and Jun Zhou[1]*

[1]Department of Radiation Oncology and Winship Cancer Institute,

Emory University, Atlanta, GA, 30322, USA

*Email: jun.zhou@emory.edu

## Abstract:

Active breath-hold techniques effectively mitigate respiratory motion but pose challenges for patients who are ineligible for the procedure. Conventional treatment planning relies on multiple energy layers, extending delivery time due to slow layer switching. We propose to use pin ridge filters (pRFs), initially developed for FLASH radiotherapy, to construct a single energy beam plan and minimize dose delivery time. The conventional ITV-based free-breathing treatment plan served as the reference. A GTV-based IMPT-DS plan with a downstream energy modulation strategy was developed based on a new beam model that was commissioned using the maximum energy of the IMPT plan. Consequently, a nested iterative pencil-beam-direction (PBD) spot reduction process removed low-weighted spots along each PBD, generating pRFs with coarser resolution. Finally, the IMPT-DS plan was then converted into an IMPT-pRF plan, using a monoenergetic beam with optimized spot positions and weights. This approach was validated on lung and liver SBRT cases (10 Gy RBE ×5). For the lung case, the mean lung-GTV dose decreased from 10.3 Gy to 6.9 Gy, with delivery time reduced from 188.79 to 36.16 seconds. The largest time reduction was at 150°, from 47.4 to 3.99 seconds. For the liver case, the mean liver-GTV dose decreased from 5.7 Gy to 3.8 Gy, with delivery time reduced from 111.13 to 30.54 seconds. The largest time reduction was at 180°, from 38.57 to 3.94 seconds. This method significantly reduces dose delivery time and organ-at-risk dose. Further analysis is needed to validate its clinical feasibility.

# 1. Introduction:

Intensity-Modulated Proton Therapy (IMPT) utilizes Bragg peaks to deliver a precise and conformal dose to the target while sparing organs at risk (OAR). However, IMPT is highly sensitive to respiratory motion due to the interplay between spot scanning and tumor movement in the abdominal or thoracic regions. This interaction can lead to dose heterogeneities, particularly when the target moves perpendicular to the beam, causing misalignment between delivered spots and the tumor position.

To mitigate these effects, various motion management strategies have been developed, including rescanning, gating, tumor tracking, and breath-hold (BH) techniques[1,2]. Among these, BH techniques such as deep inspiration breath-hold (DIBH) and active breath control (ABC) are considered the most effective, as they immobilize the tumor during beam delivery, ensuring a repeatable lung volume and tumor position for improved dosimetric accuracy[3,4]. However, BH has practical limitations, particularly for patients with compromised lung function who may struggle to maintain a stable breath-hold, making them ineligible for the technique or leading to treatment inconsistencies[5]. Additionally, BH techniques can prolong treatment time due to patient coaching and frequent interruptions, increasing discomfort and reducing overall efficiency[6].

For patients eligible for breath-hold (BH) techniques, respiratory motion introduces an uncertainty of approximately 5 mm, with an overall standard deviation of 0.48 cm[7]. This high uncertainty, combined with the discomfort experienced during prolonged treatments, underscores the need for faster dose delivery methods that reduce the requirement for longer BH without increasing treatment margins or worsening uncertainties. In proton therapy, dose delivery is slowed by reduced dose rates at low energies and prolonged energy layer switching (ELS) times, making it impractical for many patients to sustain a breath-hold for extended durations[8]. To address these challenges, we propose a method to reduce the total proton dose delivery time, improving motion management for patients who cannot utilize current BH techniques. Faster delivery not only enhances patient comfort by reducing table time but also increases the efficiency of multi-room proton therapy systems, where all treatment rooms share the same beamline, ultimately improving overall facility throughput.

In most commercial proton therapy systems, ELS times range from 0.2 to 2 seconds, with the fastest commercially available switching times approaching 100 ms[9-11]. A promising approach to significantly reducing proton beam delivery time is the use of ridge filters (RF), which broaden the Bragg peak width, thereby reducing the required number of energy layers[12-18].

While RFs have been primarily studied in preclinical FLASH proton therapy, an active area of research due to its potential biological sparing effects. Previous studies by our group have demonstrated the feasibility of RF-modulated FLASH therapy[19]. In this paper, however, we propose a novel application of RFs to conventional proton therapy to reduce treatment time. By using the Bragg peak broadening effect of RFs, our approach aims to remove energy layer transitions without compromising dose conformity, thereby enhancing clinical efficiency.

## 2. Method:

This study employed a monoenergetic proton beam set to the maximum energy level used in conventional IMPT planning. We designed patient-specific IMPT-pRF (Intensity-Modulated Proton Therapy with pin ridge filters) plans using an inverse planning framework integrated into our clinical treatment planning system (TPS), RayStation[20], to design a pin RF plan specific for each patient. The translation of a multi-energy proton beam plan into a single-energy IMPT-pRF plan involves four steps, as illustrated in Figure 1. This schematic outlines the stages of IMPT-pRF plan creation, including the inverse planning framework's role in iteratively refining pencil beam directions (PBDs) and energy layer configurations.

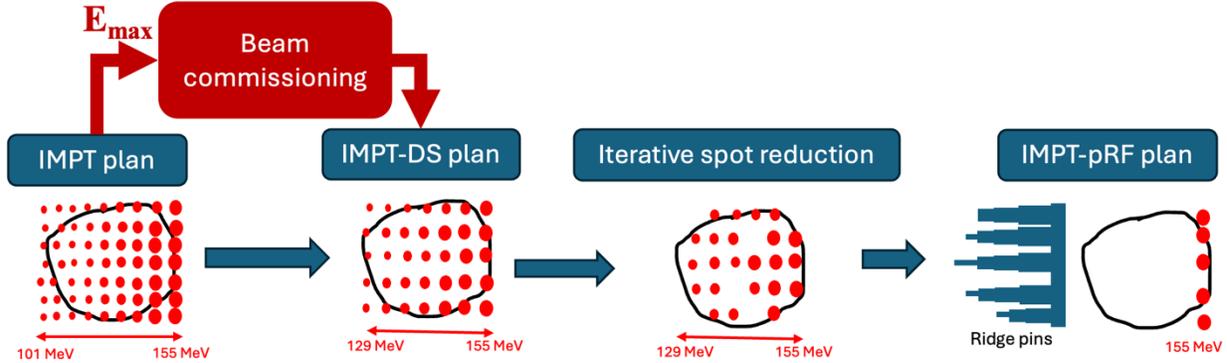

Figure 1. Schematic diagram of the workflow. The process begins with a conventional IMPT plan based on multiple energy layers. The maximum energy value from the IMPT plan is used as a benchmark to commission a beam tailored for this patient. The IMPT-DS plan is constructed using the newly commissioned beam and the original IMPT plan. An iterative spot reduction algorithm is applied in the IMPT-DS plan to remove low-weight spots and PBDs, redistributing the weights of removed spots to adjacent spots. Finally, the IMPT-DS plan is converted into a monoenergetic proton beam plan, constructed using the same commissioned beam.

## 2.1 Beam commissioning:

In conventional IMPT, the maximum energy layer corresponds to the deepest penetration depth required to cover the distal edge of the target. Selecting the maximum energy ($E_{max}$) of the conventional IMPT plan as the single energy for the IMPT-DS (Downstream Energy Modulation) plan strategy ensures penetration to the distal edge of the target while enabling a less energy layer plan. This $E_{max}$ guarantees that the Bragg peak reaches the deepest margin of the target, avoiding underdosing at the distal boundary. In order to use this single energy ($E_{max}$) for the IMPT-DS plan, we propose the commissioning of a customized beam for each patient based on $E_{max}$ of the conventional IMPT plan. The commission process involved systematic measurements of in-air lateral profiles and integrated depth dose (IDD) for $E_{max}$ (MeV) at distances of 20, 10, 5, 0, and -5 cm from the isocenter in a water phantom. The measurements were complemented by additional data collection of the in-air lateral profiles and IDDs for sub-$E_{max}$ energies at the same spatial positions to support the IMPT-DS planning framework. Figure 2 illustrates the input (measured) and simulated (Monte Carlo) IDD of the energies of one of the beams commissioned in the TPS.

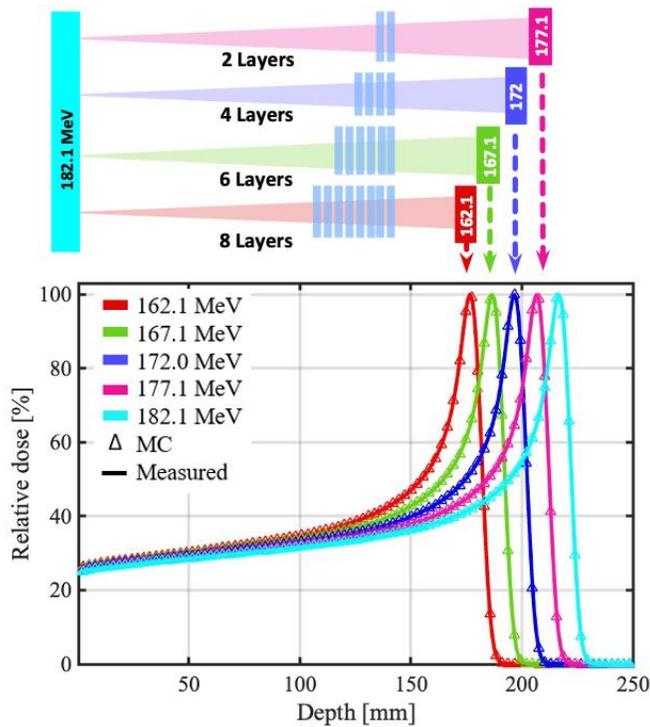

Figure 2. Comparisons of the measured and MC simulated. (a) the measured (input) and Monte Carlo (MC) simulated IDDs for 5 energies

## 2.2 IMPT-DS planning:

An important step in the process of IMPT-pRF planning is the generating of a multi-energy reference plan by using the IMPT-DS planning framework. Unlike the conventional IMPT plan, which relies on upstream energy degradation and selection. The IMPT-DS method leverages the commissioned beam model, to modulate proton energy downstream. This model incorporates a commissioned $E_{max}$ energy beam, enabling precise depth-dose control without physical energy adjustments. Similar to the conventional IMPT plan, the IMPT-DS plan optimizes spot energies and weights by solving a constrained objective function. By integrating the commissioned $E_{max}$ beam model into the optimization, the IMPT-DS framework retains the inherent beam divergence caused by multiple Coulomb scattering (MCS), ensuring that the optimized spot energies and weights can be replicated in practice using a single-energy beam modulated by the translated pin-RF. This approach maintains coverage of GTV and normal tissue sparing as compared to the use of ITV in the conventional IMPT plan for motion management.

## 2.3 Iterative spot reduction:

An iterative spot reduction algorithm was employed to minimize the number of PBDs and spots per PBD to generate simplified pRFs with reduced steps and pins. The process employs a nested optimization strategy. An initial IMPT-DS plan with optimized spot weights is first generated. The inner loop iteratively removes low-weighted spots in each PBD to increase effective step widths ($\bar{s}_{p,i}$) in translated pins, achieved by redistributing weights to adjacent spots.

$$\bar{s}_{p,i} = s_{p,i+1} - s_{p,i} \tag{1}$$

Effective step widths are gradually enlarged using ascending thresholds ($l_1 < l_2 < ... < l_K$), with the removal of spots below each threshold and weights re-optimized. The outer loop removes the entire PBDs with the total sum of spot weights below a predefined threshold ($W_T$), further reducing pin counts. After completion of optimization, residual spots, and PBDs meet minimum step width ($\geq l_K$) and weight ($\geq W_T$) criteria, enhancing deliverability by ensuring sufficient MUs for high beam current delivery. This nested-loop approach balances the width and number of pins in the pRF plan and prioritizes a clinically practical IMPT-pRF configuration.

## 3.3 IMPT-pRF plan:

Single-energy IMPT using pRFs involves translating spot weights from an IMPT-DS plan into pin-RF parameters by using equations (2)-(4). Each pRF aligns with a projected PBD from the IMPT-DS plan, modulating an $E_{max}$ monoenergetic beam by varying step thicknesses and widths (derived from equation (2)) to generate a spread-out Bragg peak (SOBP).
The pRF utilizes the pyramid-shaped ridge pins mounted on a base to modulate the water equivalent depths (WET) and fluences of a monoenergetic proton PBD. This is achieved through steps of varying thicknesses and surface areas, which alter the Bragg peak (BP) energies. The resulting depth dose curve along the PBD is a weighted combination of Bragg curves corresponding to different BP depths. For each $i^{th}$ step at a $p^{th}$ position, the WET ($t_{p,i}$) determines the $i^{th}$ largest BP range ($R_{p,i}$), defined by the 80% dose falloff:

$$R_{p,i} = R_o - t_{p,i} \tag{2}$$

Key parameters include a pin base thickness of 5 mm, pin width ($S_{p,i} = 0.6\ cm$), and spot spacing (0.7 cm at isocenter), optimized to balance beam modulation quality. The relative weight $\overline{w}_{p,i-1}$, representing the proportion of protons passing through step i, is determined recursively by equation (4).

$$\overline{w}_{p,i-1} = \frac{S^2_{p,i+1} - S^2_{p,i}}{S^2_{p,i}} \tag{3}$$

$$\overline{w}_{p,i-1} = \frac{w_{p,i}}{\sum_{i=1}^{N} w_{p,i}} \tag{4}$$

The IMPT-DS planning parameters, including spot spacing and minimum spot weight summation ($W_T$= 300 MU), are linked to pRF design to ensure sparsity and deliverability. Step widths are calculated by equation (3), rounded to 1 mm resolution, and constrained by effective step width limits ($L$ = 0.1–1 mm) during PBD reduction.

## 3.4 Patient Study:

The proposed approach was validated on two stereotactic body radiation therapy (SBRT) cases at our institution, a lung, and a liver case. Each case is prescribed a total dose of 50 Gy delivered over 5 fractions in accordance with the BR-001 protocol. The objective for ITV coverage was set as 95% receiving 100% of the prescription dose, D95% = 50 Gy. Robust optimization was implemented in IMPT-DS planning with the setup and range uncertainties of 5 mm and 5%, respectively in the lung SBRT case and 5 mm and 3.5%, respectively in the liver SBRT case. These parameters are standard in our lung case protocols. The beam commissioned on the base of $E_{max}$ of the conventional IMPT plan was 152 MeV and 174 MeV for liver and lung cases respectively. The IMPT-pRF plans translated from IMPT-DS plans were designed in the GTV instead of ITV to reduce the margins introduced in the original plan due to respiratory motion. A comparison of the relative total volume receiving the prescription dose ($R_{100}=V_{Rx}/V_{GTV+5mm}$) and the $D_{mean}$ of lung-GTV and liver-GTV between the conventional IMPT and IMPT-pRF plans was done to test the proposed approach. Additionally, dosimetric parameters of nearby organs at risk (OARs) were also compared between the two plans to validate the reduced toxicity. Finally, for both treatment plans, the total dose delivery time was determined by summing three components: individual spot irradiation time, intervals between spot transitions, and energy layer switching times.

## 3. Results:

The outcomes of converting the IMPT-DS plan into pRFs and single-energy configurations for the IMPT-pRF plan for both cases are demonstrated in Figure 3. This shows a comparison between the conventional IMPT plan and IMPT-pRF dose distribution of a four-beam SBRT lung case of this study. The conventional IMPT plan was designed on the ITV while the IMPT-pRF plan was based on GTV. A comparison of Figure 1(a) with Figure 1(b), demonstrated a relatively smaller dose cloud on the target region. Since the IMPT-pRF plan is a single energy plan, and the whole process of designing the IMPT-DS plan involves reducing the spots and PBDs, the final plan has a much smaller number of spots in each beam. In the conventional four-beam IMPT plan, the beam at a gantry angle of 150° had 1863 spots, with a total dose delivery time of 47.43 seconds. The beam positioned at a gantry angle of 300° with 2642 spots, took 52.01 seconds for dose delivery. Similarly, the beam at a gantry angle of 205° comprised 1987 spots, with a dose delivery time of 46.46 seconds. Finally, the fourth beam, with a gantry angle of 180°, contained 1913 spots and delivered the dose in 42.89 seconds. In contrast, the IMPT-pRF plan significantly reduced the number of spots for the same beam angles. The beam at 150° contained only 21 spots, with a dose delivery time reduced to 3.99 seconds. The beam at 300° had 44 spots, taking 14.09 seconds, and the beam at 205° included 37 spots, delivering the dose in 10.51 seconds, while the beam at 180° had 30 spots, completing dose delivery in 7.57 seconds. Figure 3(c) shows the dose volume histogram (DVH) of GTV, and nearby organs at risk as a comparison between conventional IMPT and IMPT-pRF plan.

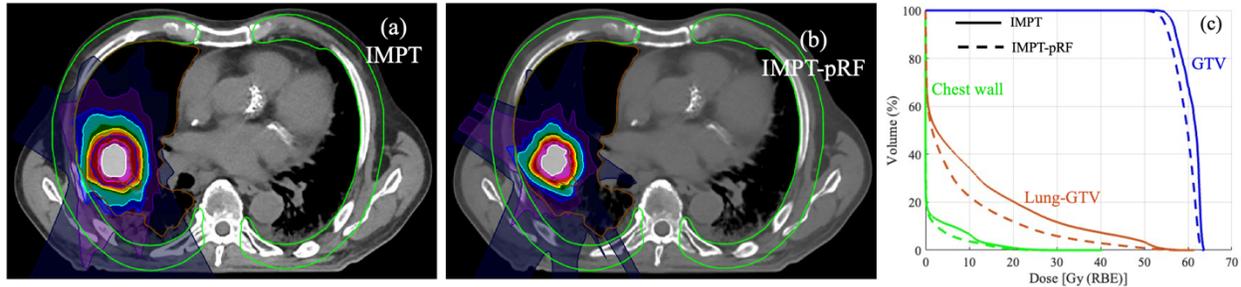

Figure 3. Lung SBRT: A comparison of the four-beam IMPT plan with the single energy proton plan using designed Ridge Filters (a) Transversal view of 2D dose distribution of IMPT plan showing the target (ITV), outlined with blue line and ROIs, lung-GTV as orange line, and chest wall as green line. (b) The 2D dose distribution of the IMPT-pRF plan, designed based on the pRF targeting the GTV, demonstrates a reduced dose cloud with improved dose conformity. (c) Dose-volume histogram (DVH) comparison for IMPT plan and IMP-pRF plans expressed as a percentage of the prescribed dose of 50 Gy in 5 fractions.

A comparison of dosimetric quantities between the two plans demonstrated much better coverage to the region of GTV with a 5 mm margin. $R_{100}$ of the IMPT-pRF plan was 1.39 compared to 0.68 from the conventional IMPT plan. Additionally, the $D_{mean}$ of lung-GTV was 10.3 Gy for the IMPT plan versus 6.9 Gy in the IMPT-pRF plan. A similar analysis was performed on a three-beam liver SBRT case, as shown in Figure 4. Consistent with the lung case, the dose cloud appears smaller in size when comparing Figure 4(a) to Figure 4(b).

Table 1. Comparison of the number of spots and beam delivery time per beam in IMPT and IMPT-pRF Plans for Both Cases analyzed in this study.

| Lung SBRT | | | | | Liver SBRT | | | | |
|---|---|---|---|---|---|---|---|---|---|
| | IMPT | | IMPT-pRF | | | IMPT | | IMPT-pRF | |
| Gantry angle | Number of spots | Time (s) | Number of spots | Time (s) | Gantry angle | Number of spots | Time (s) | Number of spots | Time (s) |
| 150° | 1863 | 47.43 | 21 | 3.99 | 150° | 1396 | 35.47 | 57 | 11.43 |
| 300° | 2642 | 52.01 | 44 | 14.09 | 290° | 1452 | 37.09 | 50 | 15.17 |
| 205° | 1987 | 46.46 | 37 | 10.51 | 180° | 1623 | 38.57 | 36 | 3.94 |
| 180° | 1913 | 42.89 | 30 | 7.57 | | | | | |

In the IMPT plan, the three beams were positioned at gantry angles of 150°, 290°, and 180° with 1396, 1492, and 1623 spots, with corresponding delivery times of 35.37, 37.09, and 38.57 seconds, respectively. In contrast, the IMPT-pRF plan, with the same gantry angles, and significantly fewer spots 57, 50, and 36 resulted in substantially reduced delivery times of 11.43, 15.17, and 3.94 seconds, respectively. Furthermore, $R_{100}$ of the IMPT plan was 2.11 and it was reduced to 0.94 in the IMPT-pRF plan. Further analysis of the dose-volume histogram (DVH) in Figure 4(c) revealed a reduction in the liver-GTV $D_{mean}$, dropping from 5.7 Gy in the IMPT plan to 3.8 Gy in the IMPT-pRF plan.

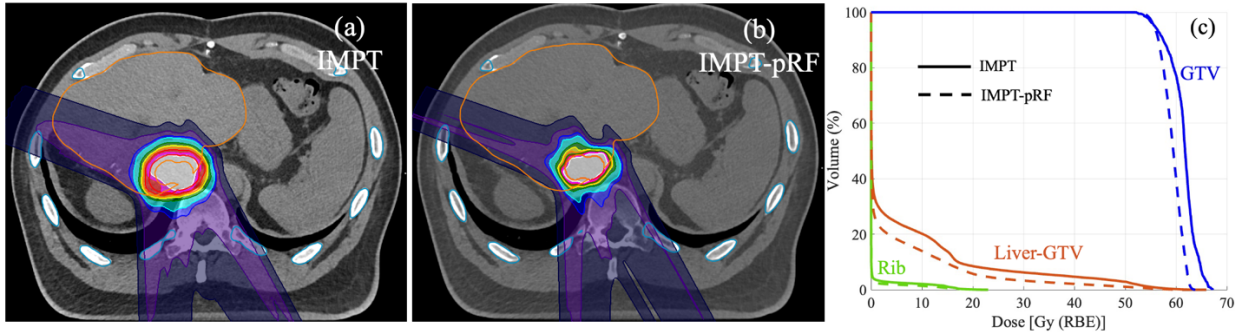

Figure 4. Liver SBRT: A comparison of the three-beam IMPT plan with the single energy proton plan using designed Ridge Filters (a) Transversal view of 2D dose distribution of IMPT plan showing the target (ITV), outlined with red line and regions of interest (ROIs), Liver-GTV as orange line, and Rib as blue line. (b) The 2D dose distribution of the IMPT-pRF plan, designed based on the pRF targeting the GTV, demonstrates a reduced dose cloud with improved dose conformity. (c) Dose-volume histogram (DVH) comparison for IMPT plan and IMP-pRF plans expressed as a percentage of the prescribed dose of 50 Gy in 5 fractions.

## 4. Discussion:

This study investigated a potentially efficient dose delivery mechanism, for patients ineligible for breath-hold, by employing a monoenergetic proton beam plan with an optimized pRF and a marked decline in the number of spots per beam. To evaluate clinical applicability, conventional IMPT plans were designed for both lung and liver cases using ITVs with respiratory motion margins. By testing our approach across two distinct anatomical sites, we aimed to assess the adaptability of the proposed method to differing organ motion and dosimetric challenges.

The clinical threshold for determining a patient's suitability for BH during proton therapy at our institution is contingent upon their ability to sustain a BH for at least 25 seconds. Prolonged delivery durations in conventional IMPT often result in extended BH cycles, increasing physical strain and potentially reducing patient compliance. In contrast, the IMPT-pRF single-energy proton model demonstrates comparable target coverage (GTV) with better normal tissue sparing to conventional IMPT while significantly reducing delivery times. In the lung SBRT case, the delivery durations were curtailed to 14.07 seconds at a 300° gantry angle, with the shortest delivery time (3.99 seconds) observed at a 150° gantry angle (see Table 1). These reductions in delivery time directly translate into fewer BH cycles, alleviating patient burden and enhancing treatment feasibility. Notably, the beam at a 300° gantry angle took the longest delivery time (44 spots, 14.09 seconds), and can be further improved by adjusting the weight threshold (WT) to selectively remove PBDs with low weights.

Additionally, the $R_{100}$ for the lung case was reduced by approximately 50%. This suggests that despite the target volume decreasing from 71.59 cc to 35 cc, the volume receiving the prescription dose of 50 Gy remained nearly equivalent. This outcome implies that the proposed method not only improves time efficiency but also maintains dosimetric quality even with smaller target volumes. A reduction in dose toxicity was also observed. As $D_{mean}$ to the lung-GTV decreased by 3.4 Gy, evidenced by the reduced dose cloud distribution in Figures 2(a) and 2(b). This reduction implies lower radiation exposure to adjacent OARs, and further mitigates the risk of complications such as radiation pneumonitis[21].

Similar results were observed in the liver SBRT case. The delivery time for the three-beam plan decreased to a maximum of 15.17 seconds and a minimum of 3.94 seconds with the IMPT-pRF approach. Notably, the beam at a gantry angle of 290° took the longest delivery time (15.17 seconds). While the number of spots in this beam was slightly lower than in the beam at 150°, the average MUs per spot for beam 290° (304.16 MUs) were 52% higher than those for beam 150°. This big difference in MUs per spot combined with only a modest difference in spot count contributed to the extended delivery time for this beam. Additionally, the average time per spot for beam 290° (301.95 ms/spot) was significantly longer than for beam 150° (198.98 ms/spot) and 180° (107.98 ms/spot). Moreover, the delivery time is also influenced by factors such as spot pattern, inter-spot distance, and total MUs of the beam. The dose delivery time for such beams can be further reduced by the adjustments in the threshold values for removing PBDs and spots from PBD during the iterative spot reduction process.

The IMPT-pRF plan targeted a GTV of 40.03 cc, compared to the conventional IMPT plan's ITV of 89.89 cc. The reduction in $R_{100}$ from 2.11 to 0.94 further supports the method's potential to maintain prescription dose coverage, aligning with findings from the lung case. Additionally, the reduced dose cloud distribution correlated with a 1.9 Gy decrease in the liver-GTV $D_{mean}$, suggesting enhanced sparing of healthy tissue. These results underscore the broader applicability of the proposed method across anatomical sites.

## 5. Conclusion:

The proposed method demonstrates a promising technique to potentially enhance proton beam delivery, achieving substantial reductions in dose delivery time by up to 75%, while preserving the plan quality. A shorter delivery time improves patient comfort by minimizing physical strain and motion-related uncertainties, particularly benefiting individuals unable to comply with breath-hold techniques. Furthermore, this approach highlights the reduced dose toxicity to adjacent OARs with lower radiation exposure persistent on both anatomical sites. These improvements not only mitigate the risk of complications but also underscore the method's ability to maintain dosimetric precision despite smaller target volumes. With pronounced efficiency and safety, this approach can potentially broaden the accessibility of radiotherapy for diverse patient populations.

**Conflicts of interest**

The authors confirm that there are no known conflicts of interest associated with this publication.

# Reference:

1	Molitoris, J. K. *et al.* Advances in the use of motion management and image guidance in radiation therapy treatment for lung cancer. *Journal of thoracic disease* **10**, S2437 (2018).
2	Pakela, J. M., Knopf, A., Dong, L., Rucinski, A. & Zou, W. Management of motion and anatomical variations in charged particle therapy: past, present, and into the future. *Frontiers in Oncology* **12**, 806153 (2022).
3	Sixel, K. E., Aznar, M. C. & Ung, Y. C. Deep inspiration breath hold to reduce irradiated heart volume in breast cancer patients. *International Journal of Radiation Oncology* Biology* Physics* **49**, 199-204 (2001).
4	Kuo, C.-C., Chang, C.-C., Cheng, H.-W. & Tsai, J.-T. Impact of Active Breathing Control-Deep Inspiration Breath Hold (ABC-DIBH) on the dose to surrounding normal structures in tangential field left breast radiotherapy. *Therapeutic Radiology and Oncology* **4** (2020).
5	Bowen, S. R. *et al.* Challenges and opportunities in patient-specific, motion-managed and PET/CT-guided radiation therapy of lung cancer: review and perspective. *Clinical and translational medicine* **1**, 1-17 (2012).
6	Oonsiri, P., Wisetrinthong, M., Chitnok, M., Saksornchai, K. & Suriyapee, S. An effective patient training for deep inspiration breath hold technique of left-sided breast on computed tomography simulation procedure at King Chulalongkorn Memorial Hospital. *Radiation Oncology Journal* **37**, 201 (2019).
7	George, R., Vedam, S. S., Chung, T. D., Ramakrishnan, V. & Keall, P. J. The application of the sinusoidal model to lung cancer patient respiratory motion. *Medical Physics* **32**, 2850-2861 (2005). https://doi.org/https://doi.org/10.1118/1.2001220
8	Zhu, M. *et al.* Effect of the initial energy layer and spot placement parameters on IMPT delivery efficiency and plan quality. *Journal of Applied Clinical Medical Physics* **24**, e13997 (2023).
9	Giovannelli, A. C. *et al.* Beam properties within the momentum acceptance of a clinical gantry beamline for proton therapy. *Medical physics* **49**, 1417-1431 (2022).
10	Saito, N. *et al.* Speed and accuracy of a beam tracking system for treatment of moving targets with scanned ion beams. *Physics in Medicine and Biology* **54**, 4849-4862 (2009). https://doi.org/10.1088/0031-9155/54/16/001
11	Pedroni, E., Meer, D., Bula, C., Safai, S. & Zenklusen, S. Pencil beam characteristics of the next-generation proton scanning gantry of PSI: design issues and initial commissioning results. *The European Physical Journal Plus* **126**, 66 (2011). https://doi.org/10.1140/epjp/i2011-11066-0
12	Courneyea, L., Beltran, C., Tseung, H. S. W. C., Yu, J. & Herman, M. G. Optimizing mini-ridge filter thickness to reduce proton treatment times in a spot-scanning synchrotron system: Optimal MRF thickness for spot-scanning. *Medical Physics* **41**, 061713 (2014). https://doi.org/10.1118/1.4876276

13	Kostjuchenko, V., Nichiporov, D. & Luckjashin, V. A compact ridge filter for spread out Bragg peak production in pulsed proton clinical beams. *Medical Physics* **28**, 1427-1430 (2001). https://doi.org/10.1118/1.1380433
14	O'Grady, F. *et al*. The use of a mini-ridge filter with cyclotron-based pencil beam scanning proton therapy. *Medical Physics* **50**, 1999-2008 (2023). https://doi.org/10.1002/mp.16254
15	Takada, Y., Kobayashi, Y., Yasuoka, K. & Terunuma, T. A miniature ripple filter for filtering a ripple found in the distal part of a proton SOBP. *Nuclear Instruments and Methods in Physics Research Section A: Accelerators, Spectrometers, Detectors and Associated Equipment* **524**, 366-373 (2004). https://doi.org/10.1016/j.nima.2004.01.069
16	Akagi, T. *et al*. Ridge filter design for proton therapy at Hyogo Ion Beam Medical Center. *Physics in Medicine and Biology* **48**, N301-N312 (2003). https://doi.org/10.1088/0031-9155/48/22/N01
17	Fujitaka, S. *et al*. Reduction of the number of stacking layers in proton uniform scanning. *Physics in Medicine and Biology* **54**, 3101-3111 (2009). https://doi.org/10.1088/0031-9155/54/10/009
18	Maradia, V. *et al*. Universal and dynamic ridge filter for pencil beam scanning particle therapy: a novel concept for ultra-fast treatment delivery. *Physics in Medicine & Biology* **67**, 225005 (2022). https://doi.org/10.1088/1361-6560/ac9d1f
19	Ma, C. *et al*. Feasibility study of modularized pin ridge filter implementation in proton FLASH planning for liver stereotactic ablative body radiotherapy. *Physics in Medicine & Biology* **69**, 245001 (2024).
20	<https://www.raysearchlabs.com/media/press-releases/2014/first-pbs--impt-proton-treatments-with-raystation/> (
21	Tsujino, K. *et al*. Predictive value of dose-volume histogram parameters for predicting radiation pneumonitis after concurrent chemoradiation for lung cancer. *International Journal of Radiation Oncology\* Biology\* Physics* **55**, 110-115 (2003).